\documentclass[showpacs,groupedaddress,assymb,amsmath]{revtex4}
\usepackage{graphicx}
\usepackage{mathrsfs}
\usepackage{amssymb}
\usepackage{amsmath}
\usepackage[raggedright]{titlesec}

\begin{document}

\title{Preparation of topological modes by Lyapunov control}

\author{Z. C. Shi$^{1}$, X. L. Zhao$^{1}$, and X. X. Yi$^{2}$\footnote{Corresponding  address:
yixx@nenu.edu.cn}}
\affiliation{$^1$ School of Physics and Optoelectronic Technology\\
Dalian University of Technology, Dalian 116024 China\\
$^2$ Center for Quantum Sciences and School of Physics, Northeast
Normal University, Changchun 130024, China}

\begin{abstract}
By Lyapunov control, we present a proposal to drive quasi-particles
into a topological  mode in  quantum systems described by a
quadratic Hamiltonian. The merit of this control is the individual
manipulations on the boundary sites.   We take the Kitaev's chain as
an illustration for  Fermi systems and show that an arbitrary
excitation mode can be steered into the Majorana zero mode by
manipulating the chemical potential of the boundary sites. For Bose
systems, taking the noninteracting Su-Schrieffer-Heeger (SSH) model
as an example, we illustrate  how to drive the system into the edge
mode. The sensitivity of the fidelity to perturbations and
uncertainties in the control fields and initial modes is also
examined. The experimental feasibility of the proposal and the
possibility to replace the continuous control field with square wave
pulses is finally discussed.
\end{abstract}

\pacs{05.30.Fk, 05.30.Jp, 02.30.Yy} \maketitle

Compared to classical computation, quantum computation
has unparallel advantages in solving   problems like factoring a
large number\cite{shor94}. However, it is difficult to realize in
practice due to decoherence caused by environments. In order to
overcome this obstacle,  topological quantum computation
\cite{bonderson08,nayak08,bonderson13,akhmerov10,MONG14} has been
proposed, where the ground states are isolated from the rest energy
spectrum by gaps, making it robust against perturbations. The
topological quantum computation can be performed by braiding
non-Abelian anyons \cite{kitaev03,kitaev06} while the evolution of
the system, protected by topology, is described by  a nontrivial
unitary transformation.  The simplest example of the non-Abelian
anyons is the Majorana fermions which are self-conjugate
quasiparticles and have been extensively studied both theoretically
and experimentally. Recently, the Majorana fermions are predicted to
exist in fractional quantum Hall system \cite{read00}, interface
between topological insulator \cite{hasan10,qi11}, topological
superconductors
\cite{fu08,sau2010,cheng10,seradjeh11,biswas13,stoudenmire11}, solid
state system \cite{alicea12}, optical lattices \cite{kraus12,mei12}
and  spin chains \cite{niu12}. Although there are great  progress in
this field,  how to prepare and manipulate   Majorana fermions in
quantum systems remains  challenging.

Generally speaking, a quantum system  cannot evolve into a desired
state  without any quantum controls \cite{alessandro07}. While most
readers are  familiar with the feedback control,  here we begin with
introducing Lyapunov-based quantum  control. The Lyapunov control
refers to the use of Lyapunov function to design control fields for
manipulating a dynamical system. In quantum mechanics, the evolution
of system is governed by the Schr\"odinger equation and the system
state can be described by a time-dependent vector. The Lyapunov
function then can be defined as  the distance between the
time-dependent vector and the target vector. Until now, most studies
of Lyapunov control focus on the analysis of largest invariant set
\cite{beauchard07,kuang08,coron09,wang10}, quantum state steering or
preparations \cite{yi09,wang09}.  In this work, we extend the
application of Lyapunov control and apply it to manipulate many-body
system, e.g., driving quasiparticles in a quantum many-body system.

To be specific, by the use of  Lyapunov control technique, we
present a method to manipulate the topological modes in both  Fermi
and Bose systems.  For a Fermi system described by the Kitaev model,
we show how to steer an arbitrary initial mode into the Majorana
zero mode by manipulating the chemical potential of the boundary
sites.  The system can be driven into a special Majorana zero mode
localized at one of the boundaries  when the initial mode is
represented only by creation or annihilation operators. For a Bose
system described by the noninteracting Su-Schrieffer-Heeger (SSH)
model, the control mechanism is similar to the Fermi system.
Nevertheless, due to the vanishing off-diagonal block (pairing
terms)  in the Hamiltonian, it is impossible to drive an arbitrary
superposition of operators with different sites into the target mode
except for two special cases, namely,  the modes can be solely
described  by creation (or annihilation) operators or by creation
and annihilation operators at  same site. An unconventional Lyapunov
technique is also explored to achieve the target mode while the
conventional Lyapunov control is not effective. The sensitivity of
the fidelity to perturbations and uncertainties in the control
fields and initial modes is also examined.   Finally, we show that
the control field can be replaced with square wave pulses, which
might make the realization of the control much easier in
experiments.

\section*{Results}\label{III}

In this part, we present the main results of this work by showing
how well the topological modes can be prepared via the Lyapunov
control. The details of calculation and simulation can be found in
METHODS. Without loss of generality, we consider a quantum system
described by quadratic Hamiltonian,
\begin{eqnarray}\label{1}
H_0=\sum_{i,j=1}^{N}A_{ij}^{0}\hat{a}_{i}^{\dag}\hat{a}_{j}+\frac{1}{2}
\sum_{i,j=1}^{N}\Big{(}B_{ij}^{0}\hat{a}_{i}^{\dag}\hat{a}_{j}^{\dag}
+B_{ij}^{0*}\hat{a}_{j}\hat{a}_{i}\Big{)},
\end{eqnarray}
where $\hat{a}_j$ and $\hat{a}_j^{\dag}$ denote the annihilation and
creation operators for fermions or bosons at the spatial position
$j$. ``$\ast$'' stands for complex conjugate. The $N\times N$ matrix
$A^{0}$ ($B^{0}$) with elements $A_{ij}^{0}$ ($B_{ij}^{0}$) should
satisfy $A^{0}=A^{0\dag}$ ($B^{0}=\varepsilon \widetilde{B}^{0}$) to
guarantee the hermicity of $H_0$, where   ``$\sim$'' denotes
transposition, and $\varepsilon=-1$ for fermions while
$\varepsilon=1$ for bosons. Since the commutation relations
of fermions are different  from bosons, we will study the control
for the Fermi and Boson systems separately.

\textbf{Fermi system}. We take the 1D Kitaev's chain of spinless
fermions \cite{kitaev01} as an example. The Hamiltonian reads,
\begin{eqnarray}\label{15}
H_{0}^{f}&=&\sum_{j=1}^{N-1}\Big{(}-J\hat{a}_{j}^{\dag}\hat{a}_{j+1}+\Delta
\hat{a}_{j}\hat{a}_{j+1}+H.c.\Big{)}+\sum_{j=1}^{N}\mu
\hat{a}_{j}^{\dag}\hat{a}_j,
\end{eqnarray}
where $J$ and $\Delta$ are hopping and pairing amplitude,
respectively. $a_j$ ($a_j^{\dag}$) is the fermionic annihilation
(creation) operation at site $j$, and $\mu$ represents the chemical
potential. By the pioneering work \cite{kitaev01}, one can find that
there exist two different topological phases when parameters change.
The quantum critical line separating those phases is given by
$2|J|=|\mu|$ and $\Delta=0$. To be specific, the parameter
satisfying  $2|J|>|\mu|$ and $\Delta\neq0$ is a nontrivial
topological phase which can support a Majorana zero mode at the
boundaries. In following, we set $\Delta=1$ and $J=\mu=2$ to ensure
the existence of  the Majorana zero mode in the Kitaev's chain. The
Majorana zero mode can be revealed  by solving the secular equation
of the BdG Hamiltonian,
\begin{eqnarray}\label{15}
\left[ {\begin{array}{*{20}{c}}
   A^{0} & B^{0}  \\
   {{-B^{0*} }} & { - {A^{0*} }}  \\
\end{array}} \right]\left[ {\begin{array}{*{20}{c}}
   {X}^{i}  \\
   {Y}^{i}  \\
\end{array}} \right]&=&\varepsilon_i\left[ {\begin{array}{*{20}{c}}
   {X}^{i}  \\
   {Y}^{i}  \\
\end{array}} \right], ~~i=1,...,2N,
\end{eqnarray}
where  the  elements of matrices $A^{0}$ and $B^{0}$ are
\begin{eqnarray}\label{15}
A_{n,j}^{0}&=&\mu\delta_{nj}-J(\delta_{n,j+1}+\delta_{n,j-1}),   \nonumber\\
B_{n,j}^{0}&=&2\Delta(\delta_{n,j-1}-\delta_{n,j+1}).     \nonumber
\end{eqnarray}
It can be found easily that ${X}^{i}={X}^{i*}$ and
${Y}^{i}={Y}^{i*}$ due to the time-reversal symmetry of the
Hamiltonian.

Fig. 1(a) demonstrates the eigenvalues of the BdG Hamiltonian, while
Fig. 1(b)-(c) and Fig. 1(d)-(e) show the distribution of the left
and right Majorana zero mode, respectively. As seen in this figure,
the Majorana zero mode is located near the two boundary sites of the
chain. Taking a chain of length $N=30$ for concreteness, we show in
the following that the Majorana zero mode can be achieved by
controlling the chemical potential at the two ends of the Kitaev's
chain. Consider two control Hamiltonians
$H_{1}^{f}=\hat{a}_1^{\dag}\hat{a}_1$ and
$H_{2}^{f}=\hat{a}_{N}^{\dag}\hat{a}_{N}$,  the nonzero elements of
matrices  $A^{k}$ given by Eq.(\ref{20})  corresponding to
the control Hamiltonian $\mathcal{H}_k^{f}$ $(k=1,2)$ are
$A_{1,1}^{1}=1$ and $A_{N,N}^{2}=1$.

Suppose that the initial mode is an equally weighted  superposition
of all sites, namely the initial mode can  be expressed as
$\mathfrak{\hat{a}}(0)=\sum_{j=1}^{N}C_{j}(0)\hat{a}_{j}+D_{j}(0)\hat{a}_{j}^{\dag}$
with $C_{j}(0)=D_{j}(0)=1/\sqrt{2N}$. The form of Lyapunov function
could be chosen as $V=Q^{\dag}PQ$ and the hermitian matrix $P$ could
be constructed in the following manner (see methods),
\begin{eqnarray}\label{15}
P=\sum_{i=1}^{N}p_{i}U^{i}U^{i\dag}+p_{T}U^{T}U^{T\dag}, ~~~ i\neq
T,~T=N+1.
\end{eqnarray}
Here $p_i=0$, $p_T=-1$, and $U^{T}$ is the target eigenvector. Then
the control field becomes $f_k(t)=iF_k\cdot
Q^{\dag}[P,\mathcal{H}_{k}^{f}]Q$ and we choose  $F_k=10$  for the
numerical calculations. Fig. 2 shows   the occupations of the left
and right mode as a function of evolution time, where the occupation
is defined by $O_l=|Q^{\dag}U^{30}|^2$ for the left mode,  and
$O_r=|Q^{\dag}U^{31}|^2$ for the right mode. We  observe that the
initial mode  asymptotically converges to the Majorana zero mode
with time, and the control fields  almost vanish when the
system arrives at the target mode. Further simulations show  that
this proposal works for almost arbitrary initial modes. For example,
it can also be driven to the Majorana zero mode when the initial
modes are
$\mathfrak{\hat{a}}(0)=\sin{\theta}\hat{a}_{j}+\cos{\theta}\hat{a}_{j}^{\dag}$
with $\theta\in[0,2\pi]$.

For a finite length $N$ of the Kitaev's chain, there exists a weak
interaction between the left and right mode with the interaction
strength $\lambda\propto e^{-N/\xi}$ \cite{kitaev01}, where $\xi$ is
the coherence length. Obviously, the left and right modes are
degenerate when $N/\xi\gg1$. Therefore, it is impossible to drive an
initial mode into one of the Majorana zero mode individually, if the
initial mode includes both the creation and annihilation operators
at the same site. However, when the initial mode can be represented
by
$\mathfrak{\hat{a}}(0)=\sum_{j=1}^{N}C_{j}(0)\hat{a}_{j}+D_{j}(0)\hat{a}_{j}^{\dag}$
with  constraint that $D_{j}(0)=0$ if $C_{j}(0)\neq0$ or
$C_{j}(0)=0$ if $D_{j}(0)\neq0$, it might be possible to drive the
initial mode into one of the Majorana zero mode. Fig. 3 shows this
possibility for driving the system into the right mode while the
initial mode is $\mathfrak{a}(0)=\sum_{j=1}^{N}C_{j}(0)\hat{a}_{j}$
with $C_{j}(0)=1/\sqrt{N}$. As expected, it converges to the right
mode asymptotically.

\textbf{Bose system}. For the case of bosons, we take the
noninteracting Su-Schrieffer-Heeger (SSH) model \cite{heeger88} to
show the control performance. The Hamiltonian reads
\begin{eqnarray}\label{15}
H_{0}^{b}&=&\sum_{j=1}^{N-1}\Big{\{}-J[1+\epsilon(-1)^{j}](\hat{a}_{j}^{\dag}\hat{a}_{j+1}
+H.c.)\Big{\}}+\sum_{j=1}^{N}\mu \hat{a}_{j}^{\dag}\hat{a}_j,
\end{eqnarray}
where $\epsilon$ is a parameter to change the hoping amplitude $J$,
$0\leq\epsilon\leq1$, and $\mu$ is the chemical potential. This
model can be applied to describe bosons hopping in a double-well 1D
optical lattice \cite{strabley06}. The edge mode in the topological
band has been shown in Ref. 31, which can be witnessed by the
nontrivial Zak phase \cite{zak89} of the bulk bands. Thereby  it can
be taken as the target mode in this control system, and we choose
the parameters $J=1,\mu=2$, $N=21$, and $\epsilon=0.3$ for the
following numerical calculation. Firstly, we present the results of
exact diagonalization of $\tau_z\cdot\mathbb{H}_{0}^{b}$
\cite{blaizot} in Fig. 4(a) and give the coefficients of the edge
mode in Fig. 4(b)-(e). It can be found that the edge mode is located
near the first site  of the chain, this suggests us to regulate the
on-site chemical potential (energy) of site 1 to manipulate the
system. Namely, the control Hamiltonian is suggested to be
$H_{1}^{b}=\hat{a}^{\dag}_{1}\hat{a}_{1}.$ As the Hamiltonian   is
block diagonal, we could drive the system from  an arbitrary initial
mode  to the  target mode for two special cases listed below.

\emph{\textbf{Case 1}}: The initial mode is described by an
arbitrary superposition of creation operators or annihilation
operators only. Since the annihilation and creation operators that
describe  quasi-particle modes are decoupled each other, the control
system can only converge to the annihilation or creation operators
in the target mode, respectively. For the numerical calculations, we
choose the initial mode described by a superposition of creation
operators
$\mathfrak{\hat{a}}(0)=\sum_{j=1}^{N}D_{j}(0)\hat{a}_{j}^{\dag}$
with $D_{j}(0)=1/\sqrt{N}$. That is, the initial mode contains the
creation operators of all sites in this control system. The Lyapunov
function is taken as $V=1-|Q^{\dag}Q_T|^2$ with
$\widetilde{Q}_T=\widetilde{U}^{T}=(\widetilde{X}^{32},\widetilde{Y}^{32})$,
$T=32$, and the control field is given by
$f_{1}(t)=F_1\cdot\textrm{Im}\Big{(}Q^{\dag}\mathcal{H}_{1}^{b}Q_{T}Q_{T}^{\dag}Q\Big{)}$
with $F_1=2$, where $\textrm{Im}(\cdot)$ denotes the imaginary part
of $(\cdot)$.

Fig. 5 shows the occupation of right mode as a function of evolution
time $t$. It demonstrates that the operator $\mathfrak{\hat{a}}(t)$ does
not completely converge to the right mode since the occupation of the
right mode approaches 0.5814. On the other hand, when resolving the
characteristic spectrum of the free and control Hamiltonian, one can
find that the target mode is controllable  for an arbitrary
superposition of creation operators. Next, we adopt an implicit
Lyapunov-based method to steer an arbitrary initial mode into the
right mode \cite{beauchard07}, where the Lyapunov function is
redefined as
\begin{eqnarray}\label{15}
V=1-|Q^{\dag}W_{T, \eta(Q)}|^2.
\end{eqnarray}
Here, $W_{T, \eta(Q)}$ is a target eigenvector of the matrix
$\mathcal{H}_{0}^{b}+\eta(Q)\mathcal{H}_{1}^{b}$ with
$\eta(Q)\in\mathbb{R}$ (corresponding to the right mode when
$\eta=0$, i.e., $W_{T,0}=U^{T}$). The secular equation can be
written as,
\begin{eqnarray}\label{15}
(\mathcal{H}_{0}^{b}+\eta(Q)\mathcal{H}_{1}^{b})W_{j,\eta(Q)}
=\lambda_{j,\eta(Q)}W_{j,\eta(Q)},~~~j=1,...,N,
\end{eqnarray}
where $\lambda_{j,\eta}$ represents the eigenvalues. It returns to
the secular equation of the matrix $\mathcal{H}_{0}^{b}$ when
$\eta(Q)=0$. The control field can be rewritten as
$f_{1}(t)=\eta(Q)+f_{1}^{\prime}(t)$, where $\eta(Q)$ is implicitly
defined as
\begin{eqnarray}\label{15}
\eta(Q)=\theta(1-|Q^{\dag}W_{T, \eta(Q)}|^2).
\end{eqnarray}
Here $\theta(t)$ is a slowly varying real function satisfying
 $\theta(0)=0$ and $\theta(t)>0$ for every $t>0$. We
set $\theta(t)=0.5t$ for simplicity. By taking the time derivative of
$V$, one can find
\begin{eqnarray}\label{15}
\dot{V}=F_1\cdot
f_{1}^{\prime}(t)\cdot\textrm{Im}\Big{(}Q^{\dag}\mathcal{H}_{1}^{b}W_{T,
\eta(Q)}W_{T, \eta(Q)}^{\dag}Q\Big{)},
\end{eqnarray}
where $F_1$ is an positive constant. We can choose the control field
$f_{1}^{\prime}(t)=-F_1\cdot\textrm{Im}\Big{(}Q^{\dag}\mathcal{H}_{1}^{b}W_{T,
\eta(Q)}W_{T, \eta(Q)}^{\dag}Q\Big{)}$ with $F_1=1$ to guarantee
$\dot{V}\leq0$. Fig. 6 demonstrates the dynamics of occupation of
the right mode, we find that it can reach about 0.9887 when
completing the control. Hence an arbitrary initial mode can be
steered to the right mode by making use of the implicit Lyapunov
function.

\emph{\textbf{Case 2}}: The initial mode is an arbitrary
superposition of creation and annihilation operators at the same
site only, i.e.,
\begin{eqnarray}\label{15}
\mathfrak{\hat{a}}(0)=C_{j}(0)\hat{a}_{j}+D_{j}(0)\hat{a}_{j}^{\dag}.
\end{eqnarray}
In this case, the Lyapunov function is chosen a bit different from
before, which becomes $V=2-|Q^{\dag}Q_{T1}|^2+|Q^{\dag}Q_{T2}|^2$
with $\widetilde{Q}_{T1}=(\widetilde{X}^{11},\widetilde{Y}^{11})$
and $\widetilde{Q}_{T2}=(\widetilde{X}^{32},\widetilde{Y}^{32})$.
Subsequently, the control field can be straightforwardly taken as
$f_{1}(t)=F_1\cdot\textrm{Im}\Big{(}Q^{\dag}\mathcal{H}_{1}^{b}Q_{T1}Q_{T1}^{\dag}Q
-Q^{\dag}\mathcal{H}_{1}^{b}Q_{T2}Q_{T2}^{\dag}Q\Big{)}$. We set
$C_{2}(0)=\sqrt{1.2}$, $D_{2}(0)=\sqrt{0.2}$ while the other
coefficients vanish and $F_1=1$ for numerical calculation. The
occupations of the left and right mode are given  in Fig. 7. As
expected, the Lyapunov function reaches its minimum when the system
arrives  at the edge mode. The final mode could be approximately
written as $\mathfrak{\hat{a}}(T)\approx
C_{2}(0)\hat{b}_T+D_{2}(0)\hat{b}_T^{\dag}$, showing that we have
realized the edge mode. Note that the occupation difference
$|O_r-O_l|=1$ could not guarantee that the final mode converges  to
the edge mode, which is distinct to the aforementioned  cases
$|O_r+O_l|=1$. As the evolution of the coefficients of the operator
is unitary (see equation (\ref{11})) when $B=0$, the coefficients
should satisfy
$\sum_{j=1}^{N}\big{[}|C_j(t)|^2+|D_j(t)|^2\big{]}=constant$, i.e.,
it is invariant during the evolution. From the numerical
calculation, we can find that the final mode can be approximately
written as $\mathfrak{\hat{a}}(T)\approx
C_{2}(0)\hat{b}_T+D_{2}(0)\hat{b}_T^{\dag}$, indicating that the
coefficients of the other quasiparticle modes almost vanish.

\section*{Discussions}

Until now, we have achieved the goal  of driving the initial mode of
many-body system into a desired quasi-particle mode. The proposal
needs to know exactly the system Hamiltonian and the initial mode,
as well as  to implement precisely the control fields. However, this
may be difficult in practice. In experiments, we often encounter
uncertainties in the initial modes, perturbations in the control
fields, and uncertainties in the Hamiltonian. In previous section,
the proposal has been implemented in the Fermi and Bose systems
without any perturbations or uncertainties. In following, we discuss
the effect of perturbations and uncertainties in the control fields,
initial modes and Hamiltonian on the performance of the control.

We first examine the effect of uncertainties in the initial mode and
perturbations in the control fields. Taking
$\mathfrak{\hat{a}}(0)=\hat{a}_{1}$ in the Fermi system as the
initial mode without uncertainties, we can write the initial mode
with uncertainties as
$\mathfrak{\hat{a}}^{\prime}(0)=\sqrt{1-\epsilon}
\mathfrak{\hat{a}}(0)+\sqrt{\epsilon}\mathfrak{\hat{a}}_{j}$ with
$\epsilon$ quantifying the uncertainties. The dependence of the
fidelity on $\epsilon$ is plotted in Fig. 8(a). For the control
field with perturbations, we write  it as
$f_k^{\prime}(t)=(1+\delta)f_k(t)$ with  $f_k(t)$ representing the
perturbationless control field. The dependence of the fidelity on
the perturbations is presented in Fig. 8(b).   One can find from
Fig. 8 that the fidelity is more sensitive to the uncertainties in
the initial mode, while it is robust against the perturbations in
the control fields. In fact, from the principle of the Lyapunov
control, it is suggested that the fidelity of the control process is
sensitive to the sign rather than the amplitude of the control
fields. This observation  can be used to understand the robustness
against the perturbations in the control fields.

In a more realistic circumstance, individual controls on the
boundary sites are difficult to implement, which means that the
control on the boundary sites might affect the on-site chemical
potential of their nearest neighbors. Suppose that the chemical
potential of  the nearest-neighbor sites, which is affected by the
control fields, can be characterized by
$f_k^{\prime\prime}(t)=\delta\cdot f_k(t)$, i.e., the on-site
chemical potential of 2\emph{nd} and $(N-1)$\emph{th} site are
replaced by $(1+ f_k^{\prime\prime}(t))\mu$.  The results in Fig. 9
suggest that the fidelity keeps  high even though the control fields
have influences on the nearest-neighbor sites.

On the other hand, the Lyapunov control requires to know the system
Hamiltonian exactly, which may be difficult in practice. One
then may ask how does the control performance change if there exist
uncertainties in the Hamiltonian. We now turn to study this
problem. The Hamiltonian with uncertainties can be written as
$H_0^{\prime}=H_0+\delta H_0$. Here, $\delta H_0$ denotes the
deviation (called uncertainties) of the Hamiltonian in the control
system. This deviation might manifest in  the hopping amplitude $J$,
pairing   $\Delta$, or the chemical potential $\mu$. As the control
is exerted  on the boundary sites only, we study the deviation in
the boundary sites and the bulk sites, separately. Fig. 10(a) shows
the fidelity as a function of the deviations in the boundary
Hamiltonian, $Z^{\prime}_{j}=(1+\delta)Z_{j}$ ($Z_{j}\equiv J,
\Delta, \mu$, where $j=1,N$). It finds that the deviations caused by
the boundary Hamiltonian do not have a serious impact on the
fidelity. When the deviation happens in the bulk sites, for example,
the on-site chemical potential $\mu^{\prime}_{j}$ of the bulk sites
is replaced with $\mu^{\prime}_{j}=(1+\varepsilon)\mu_{j}$ (note
that site $j$ is randomly chosen from the bulk, and $\varepsilon$ is
an random number, $\varepsilon\in[-0.02,0.02]$), we consider   $n$
($n=1,...,20$) uncertainties appearing  simultaneously at each
instance of evolution time. In other words, we simulate $n$
fluctuations for the on-site chemical potentials, where each
fluctuation is generated for  a randomly chosen site $n$, the value
of fluctuations for chosen sites is randomly created and denoted by
$\varepsilon$. By performing the extensive numerical simulations, we
demonstrate the results in Fig. 10(b). It can be found that the
quantum system is robust against small uncertainties since the
fidelity is always larger than 97.9\%. An interesting observation is
that with the number of fluctuations increasing, the fidelity
increases. This can be understood as follows. Firstly, the small
deviations  cannot close the gaps in the topological system, thus
the fidelity would not deteriorate sharply. Secondly, although more
uncertainties participate in the control procedure, the average of
the uncertainties  almost approaches zero as the average of the
random number $\varepsilon$ is zero.

Since the form of control field generally takes $f_k(t)=iF_k\cdot
Q^{\dag}[\hat{P},\mathcal{H}_{k}]Q$, the amplitude of the control
fields  may change fast with   time, which increases the difficulty
in the realizations. It is believed that the square wave pulses can
be readily achieved in experiments. Therefore we try to take the
square wave pulses instead of $f_k(t)=iF_k\cdot
Q^{\dag}[\hat{P},\mathcal{H}_{k}]Q$ for the control field. The
principle to design the square wave pulses should satisfy,
\begin{eqnarray}\label{21}
f_k(t)=\left \{
\begin{array}{rl}
    F^{\prime}_k,~~~ f_k(t)>0, \\
    -F^{\prime}_k,~~~f_k(t)<0. \\
\end{array}
\right.
\end{eqnarray}
As an example, we focus on the Bose system whose parameters are the
same as in  Fig. 7 except that the control field $f_1(t)$ is
replaced by the equation (\ref{21}) with $F^{\prime}_1=0.04$. Fig.
11 demonstrates the results for the square wave pulses of the
control field and it can also achieve the edge mode eventually. On
the other hand, we find that convergence time is shortened as well.
Of course, the square wave pulses of the control fields can also be
applied to the Fermi system.

Finally, we would like to discuss on the experimental feasibility
for the present control protocols. The SSH model can be
experimentally realized by $^{87}$Rubidium atoms \cite{bucker11} in
1D double-well optical lattice \cite{barnett13}. The implementations
of Lyapunov control require to perform operations defined by the
control Hamiltonians with strengths defined by the control fields.
In our case, the control Hamiltonians are the particle number
operators of the boundary sites, and the control can be
experimentally realized by manipulating the on-site chemical
potentials of the boundary sites. The realization of  Kitaev's chain
requires spinless fermions, which can be prepared   in an optical
lattice by trapping the fermions and the BEC reservoir with Feshbach
molecules (the couplings between them can be induced by an rf-pulse)
\cite{jiang11}. By driving the fermions with Raman laser to produce
a strong  effective coupling, the system in this situation is
equivalent to the Kitaev's chain. In order to realize the
control Hamiltonians, one can adopt additional lasers to control the
chemical potentials of the boundary sites,  where the intensity of
lasers is simulated  by square wave pulses (e.g., see $f_1(t)$ in
Fig. 11(b)). In addition, we can realize the effective Kitaev's
chain in the quantum-dot-superconductor system \cite{sau2013}, a
linear array with quantum dots linked by s-wave superconductors with
normal and anomalous hoppings. In this system, the chemical
potential in each quantum dot can be controlled individually by gate
voltages with a high degree of precision. Alternatively, the
Kitaev's chain can also be achieved in the system which consists of
a strong spin-orbit interaction semiconductor nanowire (in the low
density limit) coupling to a superconductor in magnetic field
\cite{lutchyn10,oreg10}. Then the boundary chemical potential can be
controlled by local gates \cite{alicea11,das12}. Most recently, the
observation of Majorana fermions in this system has also been
observed  in experiments \cite{das12,mourik12}.

In summary, we present a scheme to prepare quasi-particle mode by
Lyapunov control in the both Fermi and Bose systems. For the Fermi
system, we choose  the Kitaev's model as an illustration and specify
the Majorana zero mode as the target mode. The results show that by
controlling the chemical potential at the two boundary sites, the
system can be driven asymptotically into one of the Majorana zero
mode such as the right mode. In contrary, the situation for bosons
is different due to the commutation relations. As an example, in the
noninteracting SSH model, we show how to prepare the edge mode by
the control fields. In particular, we apply the implicit
Lyapunov-based technique to the boson system which provides us with
a new way to steer the bosons. The robustness of the fidelity
against perturbations and  uncertainties is also examined. Finally,
we try  to replace the control fields with square wave pulses, which
might help realize the control fields more easily in experiments
since it is difficult to apply a fast time-varying control fields in
practice.

\section*{Methods}
In this part, we give the derivation of the control scheme, starting
with the quadratic Hamiltonian,
\begin{eqnarray}\label{1}
H_0=\sum_{i,j=1}^{N}A_{ij}^{0}\hat{a}_{i}^{\dag}\hat{a}_{j}+\frac{1}{2}
\sum_{i,j=1}^{N}\Big{(}B_{ij}^{0}\hat{a}_{i}^{\dag}\hat{a}_{j}^{\dag}
+B_{ij}^{0\ast}\hat{a}_{j}\hat{a}_{i}\Big{)}.
\end{eqnarray}
For the case of fermions, we denote the Hamiltonian by $H_{0}^{f}$,
i.e., $H_{0}^{f}=H_0$. The operators obey the anticommutation
relations: $\{\hat{a}_i,\hat{a}_j^{\dag}\}=\delta_{ij},
\{\hat{a}_i,\hat{a}_j\}=0,$ and
$\{\hat{a}_i^{\dag},\hat{a}_j^{\dag}\}=0.$ Define a time-dependent
fermionic operator,
\begin{eqnarray}\label{15}
\mathfrak{\hat{a}}(t)=\sum_{j=1}^{N}C_{j}(t)\hat{a}_{j}+D_{j}(t)\hat{a}_{j}^{\dag},
\end{eqnarray}
where the operators $\hat{a}_j$ and $\hat{a}_j^{\dag}$ are time-independent
while the coefficients are time-dependent. It is easy to
check  that $\sum_{j=1}^{N}\big{[}|C_j(t)|^2+|D_j(t)|^2\big{]}=1$
according to the anti-commutation relation
$\{\mathfrak{\hat{a}}(t),\mathfrak{\hat{a}}^{\dag}(t)\}=1$. In the
Heisenberg picture, the evolution of this operator
satisfies ($\hbar=1$),
\begin{eqnarray}\label{15}
i\mathfrak{\dot{\hat{a}}}(t)=[\mathfrak{\hat{a}}(t),H_{0}^{f}].
\end{eqnarray}
After a brief algebraic operation, the equation becomes
\begin{eqnarray}\label{4}
i\mathfrak{\dot{\hat{a}}}(t)&=&\sum_{n,j=1}^{N}
\bigg{\{}\Big{[}C_j(t)A_{jn}^{0}+D_j(t)(-B_{jn}^{0\ast})\Big{]}\hat{a}_n
+\Big{[}C_j(t)B_{jn}^{0}+D_j(t)(-A_{jn}^{0\ast})\Big{]}\hat{a}_n^{\dag}\bigg{\}}.
\end{eqnarray}
The evolution of coefficients  $C_j(t)$ and $D_j(t)$ then can be
written in a compact form of matrix,
\begin{eqnarray}\label{5}
-i\dot{Q}&=&{\mathcal{H}_{0}^{f}}Q,  \nonumber\\
Q&=&\left[ {\begin{array}{*{20}{c}}
   {C}^{\ast}(t)  \\
   {D}^{\ast}(t)  \\
\end{array}} \right],
\mathcal{H}_{0}^{f}=\left[ {\begin{array}{*{20}{c}}
   A^{0} & B^{0}  \\
   {{-B^{0*} }} & { - {A^{0*} }}  \\
\end{array}} \right],
\end{eqnarray}
where
$\widetilde{C}^{\ast}(t)=\big{(}C_{1}^{\ast}(t),...,C_{N}^{\ast}(t)\big{)}$
and
$\widetilde{D}^{\ast}(t)=\big{(}D_{1}^{\ast}(t),...,D_{N}^{\ast}(t)\big{)}$.
We use the Gothic letter $\mathcal{H}_0$ to denote  the matrix in
equation (\ref{5}) corresponding  to the Hamiltonian $H_0$ in
equation (\ref{1}) for simplicity  hereafter.

For the Fermi system,  the quadratic Hamiltonian can be rewritten as
$H_{0}^{f}=\frac{1}{2}\alpha^{\dag}\mathbb{H}_{0}^{f}\alpha$ up to a
constant factor ($\mathbb{H}_{0}^{f}$ called
Bogoliubov-de-Gennes(BdG) Hamiltonian), where
$\widetilde{\alpha}=(\hat{a}_1,...,\hat{a}_N,\hat{a}_1^{\dag},...,\hat{a}_N^{\dag})$.
Clearly, $\mathbb{H}_{0}^{f}=\mathcal{H}_{0}^{f}$. In fact,
the equation (\ref{5}) is actually the BdG-Schr\"odinger
equation\cite{gennes66}, where $Q$ is the quasi-particle wave
function in the Nambu representation. One can claim that if
$\varepsilon_l$ is an eigenvalue of $\mathbb{H}_{0}^{f}$ with
corresponding eigenvector
$\widetilde{U}^{l}=({\widetilde{X}}^{l},{\widetilde{Y}}^{l})$,~$l=1,...,N$:
\begin{eqnarray}\label{15}
\mathbb{H}_{0}^{f}U^l=\varepsilon_lU^l,
\end{eqnarray}
$\widetilde{U}^{2N+1-l}=({\widetilde{X}}^{2N+1-l},{\widetilde{Y}}^{2N+1-l})$
is also an eigenvector with eigenvalue $-\varepsilon_l$, i.e.,
\begin{eqnarray}\label{15}
\mathbb{H}_{0}^{f}U^{2N+1-l}=-\varepsilon_lU^{2N+1-l},
\end{eqnarray}
where  ${X}^{l*}={Y}^{2N+1-l}$,  ${Y}^{l*}={X}^{2N+1-l}$,
$\widetilde{X}^{l}=({X}^{l}_{1},...,{X}^{l}_{N})$, and
$\widetilde{Y}^{l}=({Y}^{l}_{1},...,{Y}^{l}_{N})$. Thus the
eigenvalues come in pairs $\pm\varepsilon_l$ for the BdG Hamiltonian
$\mathbb{H}_{0}^{f}$ \cite{blaizot}. Diagonalizing   the BdG
Hamiltonian, the quasi-particles can be represented by
annihilation (creation) operators $\hat{b}_l$ ($\hat{b}_l^{\dag}$),
\begin{eqnarray}\label{15}
\hat{b}_l&=&\sum_{j=1}^{N}(X_{j}^{l}\hat{a}_j+Y^{l}_{j}\hat{a}_j^{\dag}), \nonumber\\
\hat{b}_l^{\dag}&=&\sum_{j=1}^{N}(X_{j}^{2N+1-l}\hat{a}_j+Y_{j}^{2N+1-l}\hat{a}_j^{\dag}),
\end{eqnarray}
where $l=1,...,N$. In terms of  the quasi-particle modes, the
Hamiltonian can be written as
$H_{0}^{f}=\sum_{l=1}^{N}\varepsilon_{l}\hat{b}_l^{\dag}\hat{b}_l$,
where $\varepsilon_l$ are the energy of the quasi-particle
$\hat{b}_l$.

Let one of the quasi-particle modes be the target mode which we want
to prepare, e.g., $\mathcal {\hat{T}}=u\hat{b}_T+v\hat{b}_T^{\dag}$
where $u$ and $v$ are arbitrary constants. $\hat{b}_T$ and
$\hat{b}_T^{\dag}$ are the annihilation and creation operators of
the target mode, respectively. The goal is to design control fields
that can drive any initial modes to the target one. It should be
noticed that we cannot choose the target arbitrarily because it
depends on the free Hamiltonian. In other words, we need a
stationary target mode which does not evolve  under the free
Hamiltonian.  As the edge mode is robust against perturbations, we
focus on the preparation of it.  The evolution described by the
equation (\ref{5}) is unitary since $\mathcal{H}_{0}^{f}$ is
hermitian. As a result the sum
$M=\sum_{j=1}^{N}\big{[}|C_j(t)|^2+|D_j(t)|^2\big{]}$ remains
unchanged during the time evolution.  To make the calculation clear,
we write the target mode as
$\hat{b}_{T}=\sum_{j=1}^{N}X_{j}^{T}\hat{a}_{j}+Y_{j}^{T}\hat{a}_{j}^{\dag}$,
in which
$\widetilde{U}^{T}=({\widetilde{X}}^{T},{\widetilde{Y}}^{T})$ is an
eigenvector of the BdG Hamiltonian $\mathbb{H}_{0}^{f}$, meanwhile
it is also a  solution of equation (\ref{5}). Namely,
$\widetilde{Q}_{T}=\widetilde{U}^{T}=
({\widetilde{X}}^{T},{\widetilde{Y}}^{T})
=({\widetilde{C}_{T}}^{\ast}(t),{\widetilde{D}_{T}}^{\ast}(t))$,
where $\widetilde{X}^{T}=({X}^{T}_{1},...,{X}^{T}_{N})$ and
$\widetilde{Y}^{T}=({Y}^{T}_{1},...,{Y}^{T}_{N})$. Assume
that there are $K$  control Hamiltonians $H_k^{f}$ for the system in
quadratic form:
$H_k^{f}=\sum_{i,j=1}^{N}A^{k}_{ij}\hat{a}_{i}^{\dag}\hat{a}_{j}+\frac{1}{2}
\sum_{i,j=1}^{N}\Big{(}B^{k}_{ij}\hat{a}_{i}^{\dag}\hat{a}_{j}^{\dag}
+B_{ij}^{k\ast}\hat{a}_{j}\hat{a}_{i}\Big{)}$, $k=1,...,K$.
Together with the original Hamiltonian, the equation of motion for
the coefficients in the operator $\mathfrak{\hat{a}}(t)$ becomes
\begin{eqnarray}  \label{20}
-i\dot{Q}&=&\Big{(}{\mathcal{H}_{0}^{f}
+\sum_{k=1}^{K}f_k(t)\cdot\mathcal{H}_k^{f}}\Big{)}Q, \nonumber\\
\mathcal{H}_k^{f}&=&\left[ {\begin{array}{*{20}{c}}
   A^k & B^k  \\
   {{-B^{k*} }} & { - {A^{k*} }}  \\
\end{array}} \right],
\end{eqnarray}
where $f_k(t)$ is the control field.

There are many  choices for the Lyapunov functions, for example,
$V_1=\|Q-Q_{T}\|^2$, $V_2=1-|Q^{\dag}Q_T|^2$, and $V_3=Q^{\dag}PQ$.
Here, $\|\cdot\|$ denotes the norm. Those Lyapunov functions are
nonnegative and reach the minimum when the system  arrives at the
target. Apparently, different Lyapunov functions lead to
different invariant set and different characteristics of
convergence. In following, we choose $V=Q^{\dag}PQ$ as the
Lyapunov function to show how our scheme works while the analysis
for other Lyapunov functions are similar to it. To this end, it is
instructive to deduce the first-order time derivative of the
Lyapunov function,
\begin{eqnarray}\label{15}
\dot{V}=\sum_{k=1}^{K}f_{k}(t)\cdot
iQ^{\dag}[P,\mathcal{H}_{k}^{f}]Q,
\end{eqnarray}
where we have set $[{\mathcal{H}_0^{f}},P]=0$ by properly
constructing the matrix $P$. In order to make the time derivative of
$V$ non-positive, one can design the control fields in the following
style: $f_k(t)=iF_k\cdot Q^{\dag}[\hat{P},\mathcal{H}_{k}^{f}]Q$
with $F_k>0$. Strictly speaking, the quantum system  converges  to
the invariant set determined by the La Salle's invariance principle,
equivalent to the solution $\dot{V}=0$.

Note that the  commutation relations for bosons:
$[\hat{a}_i,\hat{a}_j^{\dag}]=\delta_{ij},$ and
$[\hat{a}_i^{\dag},\hat{a}_j^{\dag}]= [\hat{a}_i,\hat{a}_j]=0,$ are
different from fermions. Keeping this difference in mind and by an
analysis similar to the  case of fermions, one can   obtain a
dynamical evolution of operator
$\mathfrak{\hat{a}}(t)=\sum_{j=1}^{N}C_{j}(t)\hat{a}_{j}+D_{j}(t)\hat{a}_{j}^{\dag}$
with
$\Big{|}\sum_{j=1}^{N}\big{[}|C_j(t)|^2-|D_j(t)|^2\big{]}\Big{|}=1$,
\begin{eqnarray}\label{11}
-i\dot{Q}={\mathcal{H}_{0}^{b}}Q, ~~~Q=\left[
{\begin{array}{*{20}{c}}
   {C}^{\ast}(t)  \\
   {D}^{\ast}(t)  \\
\end{array}} \right],
~~~\mathcal{H}_{0}^{b}=\left[ {\begin{array}{*{20}{c}}
   A^{0} & -B^{0}  \\
   {{B^{0*} }} & { - {A^{0*} }}  \\
\end{array}} \right].
\end{eqnarray}
In this case, the matrix of BdG Hamiltonian $\mathbb{H}_{0}^{b}$ is
$\left[ {\begin{array}{*{20}{c}}
   A^{0} & B^{0}  \\
   {{B^{0*} }} & { {A^{0*} }}  \\
\end{array}} \right]$.
Therefore, we can find that
$\mathcal{H}_{0}^{b}=\mathbb{H}_{0}^{b}\cdot\tau_z$, where
$\tau_z=\sigma_{z}\otimes\mathbb{I}$, $\sigma_{z}$ is Pauli matrix
and $\mathbb{I}$ is the $N\times N$ identity matrix. The dynamics of
coefficients are not unitary in general except for $B^{0}=0$. For this
special situation, the control mechanism is analogous to the case of
fermions.

\textbf{Acknowledgement} This work is supported by the National Natural Science Foundation
of China (Grant Nos. 11175032 and 61475033).

\textbf{Author Contributions}
\noindent  X. X. Yi proposed  the idea and led the study, Z. C. Shi,
X. L. Zhao., and X. X. Yi performed the analytical and numerical
calculations, Z. C. Shi and X. X. Y prepared the manuscript, all
authors reviewed the manuscript.

\textbf{Competing Interests}
The authors declare that they have no competing financial interests.

\newpage

\begin{figure}[h]
\centering
\includegraphics[scale=0.5]{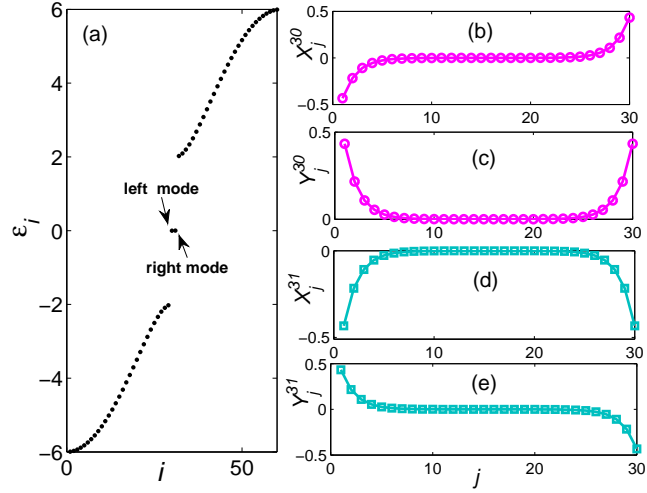}
\caption{\textbf{The energy spectrum and spatial distributions of
the BdG Hamiltonian $\mathbb{H}_{0}^{f}$ describing  the Kitaev's
chain with total number $N=30$ of sites.} We have  set the lattice
spacing as units. There exists two Majorana modes in the band gap,
i.e., the 30\emph{th} and 31\emph{th}  eigenmodes. The 30\emph{th}
eigenmode is labeled by left mode and the 31\emph{th}  is labeled by
right mode. (b) and (c) are the coefficients $X^{30}$ and $Y^{30}$
of the left mode, while (d) and (e) are the coefficients $X^{31}$
and $Y^{31}$ of the right mode.}
\end{figure}

\begin{figure}[h]
\centering
\includegraphics[scale=0.5]{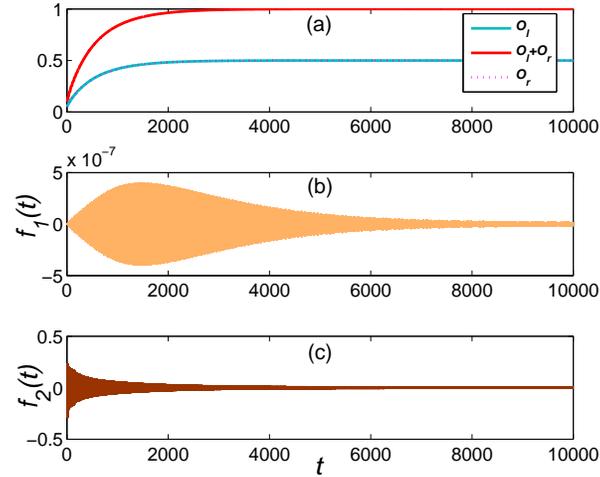}
\caption{\textbf{The dynamical evolution of system as a function of
time with initial mode
$\mathfrak{\hat{a}}(0)=\sum_{j=1}^{N}C_{j}(0)\hat{a}_{j}+D_{j}(0)\hat{a}_{j}^{\dag}$}.
$O_l$ and $O_r$ represent the occupations of the left and right
mode, while $O_l+O_r$ approaching unit implies the other
quasiparticle modes except the right and left modes are suppressed.
(b) and (c) denote the dynamical evolution of the control fields
$f_1(t)$ and $f_2(t)$, respectively.}
\end{figure}

\begin{figure}[h]
\centering
\includegraphics[scale=0.5]{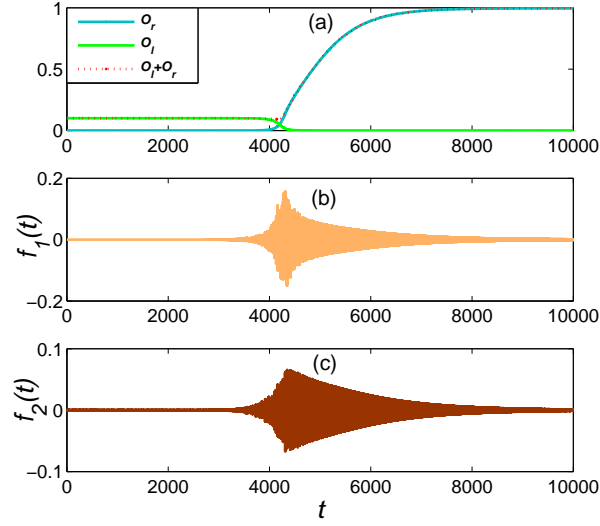}
\caption{ \textbf{The dynamical evolution of system as a function of
time with initial mode
$\mathfrak{\hat{a}}(0)=\sum_{j=1}^{N}C_{j}(0)\hat{a}_{j}$}.}
\end{figure}

\begin{figure}[h]
\centering
\includegraphics[scale=0.5]{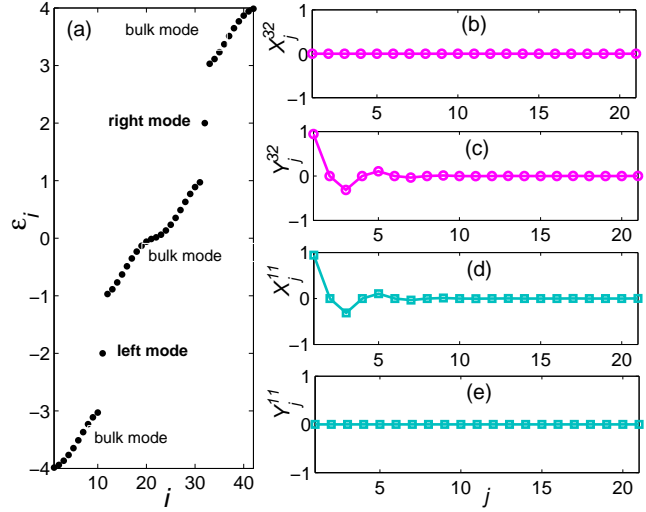}
\caption{ \textbf{The eigenvalue spectrum and spatial distributions
of the Hamiltonian $\tau_z\cdot\mathbb{H}_{0}^{b}$ in the SSH model
with $N=21$ sites.} Two edge mode are found in the band gap,
corresponding to the 11\emph{th} and 32\emph{th} eigenvectors. We
label the 11\emph{th} eigenvector as left mode while  the
32\emph{th} eigenvector is the right mode. (b) and (c) are the
coefficients $X^{11}$ and $Y^{11}$ of the left mode while (d) and
(e) are the coefficients $X^{32}$ and $Y^{32}$ of the right mode.}
\end{figure}

\begin{figure}[h]
\centering
\includegraphics[scale=0.5]{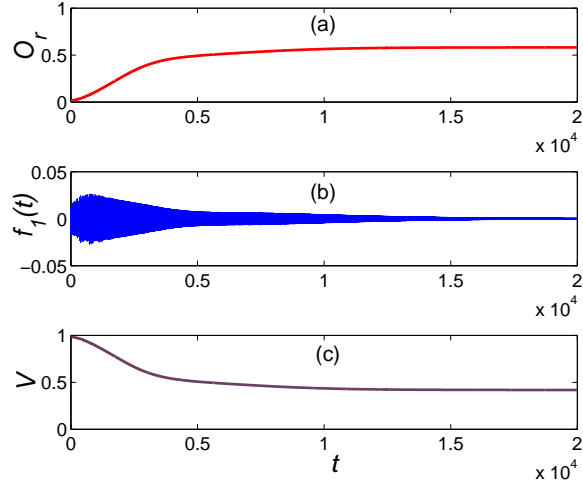}
\caption{ \textbf{The dynamical evolution of system as a function of
time with conventional Lyapunov technique and initial mode
$\mathfrak{\hat{a}}(0)=\sum_{j=1}^{N}D_{j}(0)\hat{a}_{j}^{\dag}$}.
(c) denotes the dynamical behavior of the Lyapunov function $V$.}
\end{figure}

\begin{figure}[h]
\centering
\includegraphics[scale=0.5]{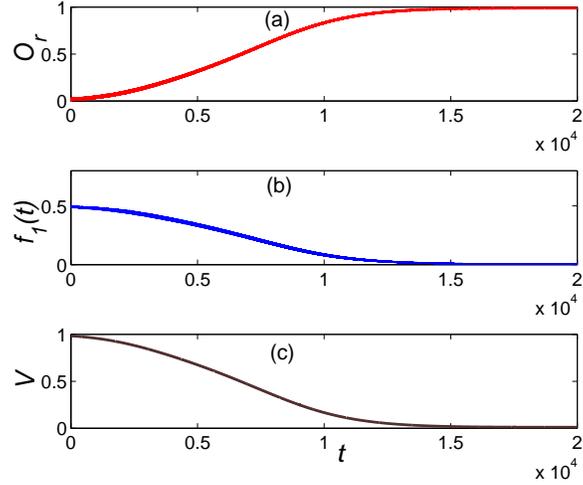}
\caption{\textbf{The dynamical evolution of system as a function of
time with unconventional Lyapunov technique}. The physical
parameters are the same to the Fig. 5 except for the Lyapunov
function.}
\end{figure}

\begin{figure}[h]
\centering
\includegraphics[scale=0.5]{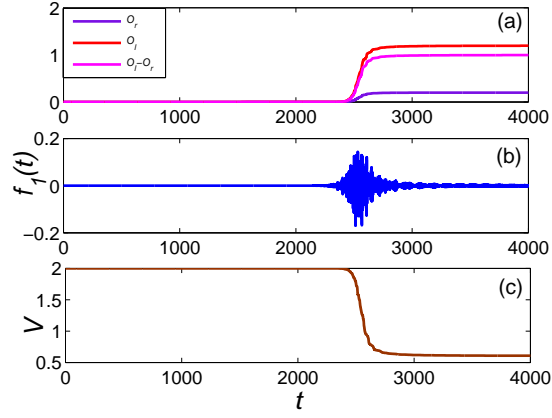}
\caption{\textbf{The dynamical evolution of system as a function of
time with the Lyapunov function
$V=2-|Q^{\dag}Q_{T1}|^2+|Q^{\dag}Q_{T2}|^2$}. It can be found that
$O_l\simeq |C_2(0)|^2$ and $O_r\simeq |D_2(0)|^2$ imply the other
quasiparticle modes being suppressed.}
\end{figure}

\begin{figure}[h]
\centering
\includegraphics[scale=0.5]{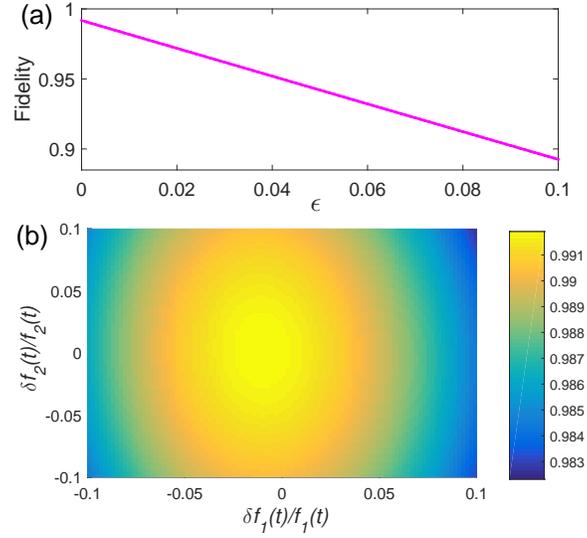}
\caption{\textbf{The fidelity versus (a) the uncertainties  in the
initial mode and (b) the perturbations  in the control fields
$f_1(t)$ and $f_2(t)$}. Other parameters are the same as in Fig.
3. The control time is terminated  when the fidelity reaches
99.15\%. One can observe that the fidelity is still above 98\% even
though there are 10\% perturbations in the control fields.}
\end{figure}

\begin{figure}[h]
\centering
\includegraphics[scale=0.5]{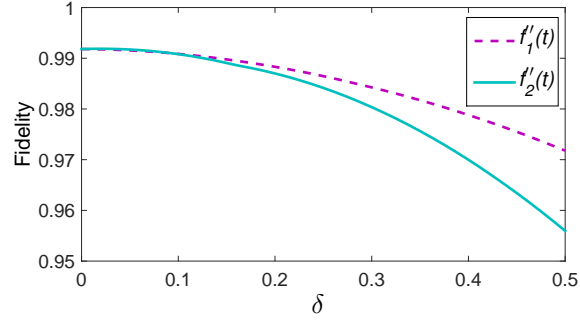}
\caption{ \textbf{The fidelity versus the chemical potential of nearest-neighbor sites
of the boundaries affected by the control fields.} We describe
this influence by   $f_k^{\prime\prime}(t)=\delta\cdot f_k(t)$.
Other parameters are the same as in Fig. 8.  $\delta=0.5$ means that
the value of control fields on the nearest-neighbor site is the half
of control fields on the boundary sites.}
\end{figure}

\begin{figure}[h]
\centering
\includegraphics[scale=0.5]{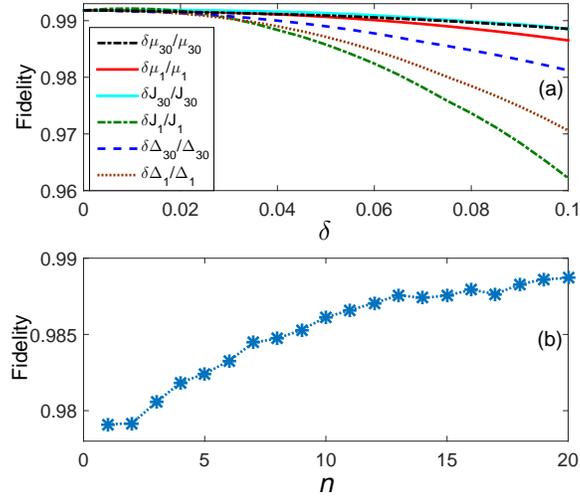}
\caption{ \textbf{The effect of uncertainties in the Hamiltonian on the fidelity.}
The influence of boundary Hamiltonian is depicted in (a).
Each point is an average over 30 simulations in (b). The horizontal
axis denotes the number of perturbations at each instance of time in
the Kitaev's chain. Other parameters are the same as in Fig. 8.
}
\end{figure}

\begin{figure}[h]
\centering
\includegraphics[scale=0.5]{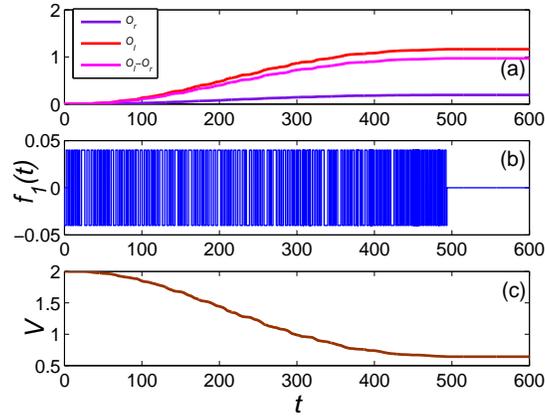}
\caption{ \textbf{The dynamical evolution of system as a function of
time with the square wave pulses}.}
\end{figure}

\end{document}